\documentclass[preprint,preprintnumbers,amsmath,amssymb]{revtex4}
\pdfoutput=1
\usepackage{epsfig}
\usepackage{graphicx}% Include figure files
\usepackage{dcolumn}% Align table columns on decimal point
\usepackage{bm}% bold math
\usepackage{threeparttable}
\usepackage{xcolor}
\usepackage[utf8]{inputenc}
\usepackage{graphicx}
\usepackage{subfigure}

\def\beq{\begin{equation}}
\def\eeq{\end{equation}}
\def\eeqn{\end{equation}}
\newcommand\iden{\leavevmode\hbox{\small1\normalsize\kern-.33em1}}

\newcommand{\bea} {\begin{eqnarray}}
\newcommand{\eea} {\end{eqnarray}}

\let\jnfont=\rm
\def\NPB#1 {{\jnfont Nucl.\ Phys.\ B }{\bf #1} }
\def\PLB#1 {{\jnfont Phys.\ Lett.\ B }{\bf #1} }
\def\EPJC#1 {{\jnfont Eur.\ Phys.\ Jour.\ C }{\bf #1} }
\def\PRD#1 {{\jnfont Phys.\ Rev.\ D }{\bf #1} }
\def\PRL#1 {{\jnfont Phys.\ Rev.\ Lett.\ }{\bf #1} }
\def\MPLA#1 {{\jnfont Mod.\ Phys.\ Lett.\ A }{\bf #1} }
\def\JPG#1 {{\jnfont J.\ Phys.\ G }{\bf #1} }
\def\CTP#1 {{\jnfont Commun.\ Theor.\ Phys.\ }{\bf #1} }
\def\JHEP#1 {{\jnfont JHEP \ }{\bf #1} }
\def\NPPS#1 {{\jnfont Nucl.\ Phys.\ Proc.\ Suppl.\ }{\bf #1} }
\def\CPC#1 {{\jnfont Comput.\ Phys.\ Commun.\ }{\bf #1} }
\def\CPL#1 {{\jnfont Chin.\ Phys.\ Lett. }{\bf #1} }
\def\APPB#1 {{\jnfont Acta\ Phys.\ Polon.\ B }{\bf #1} }

\def\lsim{\raise0.3ex\hbox{$<$\kern-0.75em\raise-1.1ex\hbox{$\sim$}}}
\def\gsim{\raise0.3ex\hbox{$>$\kern-0.75em\raise-1.1ex\hbox{$\sim$}}}
\def\PR#1 {{\jnfont Phys.\ Rept. }{\bf #1} }
\def\CHC#1 {{\jnfont Chin.\ Phys.\ C }{\bf #1} }
\def\NIMA#1 {{\jnfont Nucl.\ Instrum.\ Meth.\ A }{\bf #1} }
\def\JCAP#1 {{\jnfont JCAP \ }{\bf #1} }
\def\ASA#1 {{\jnfont Astron.\ Astrophys.\ A }{\bf #1} }  
\begin{document}

\title{\ \\[10mm] A joint explanation of $W$-mass and muon $g-2$ in 2HDM}
\author{Xiao-Fang Han$^{1}$,  Fei Wang$^{2}$, Lei Wang$^{1}$, Jin Min Yang$^{3,4}$, Yang Zhang$^{2}$}
\affiliation{$^1$ Department of Physics, Yantai University, Yantai 264005, P. R. China\\
$^2$ School of Physics and Microelectronics, Zhengzhou University, ZhengZhou 450001, P. R. China\\
$^3$ CAS Key Laboratory of Theoretical Physics, Institute of Theoretical Physics,
 Chinese Academy of Sciences, Beijing 100190,  P. R. China   \\
$^4$ School of Physical Sciences, University of Chinese Academy of Sciences, Beijing 100049,  P. R.  China}

%---------------------------------------------------------------------------

\begin{abstract}
Since both $W$-mass and muon $g-2$ can be affected by the mass splittings among extra Higgs bosons $(H,~A,~H^\pm)$ in a 2HDM, we take a model with $\mu$-$\tau$ LFV interactions to examine the two anomalies reported respectively by CDF II and FNAL. We obtain the following observations:
(i) Combined with theoretical constraints, the CDF $W$-mass measurement disfavors
$H$ or $A$ to degenerate in mass with $H^\pm$, but allows $H$ and $A$ to degenerate.
The mass splitting between $H^\pm$ and $H/A$ is required to be larger than 10 GeV.
The $m_{H^\pm}$ and $m_{A}$ are favored to be smaller than 650 GeV for $m_H<120$ GeV, and allowed to
have more large values with increasing of $m_H$.
 (ii) After imposing other relevant experimental constraints, there are parameter spaces that  simultaneously satisfy (at $2\sigma$ level) the CDF $W$-mass, the FNAL muon $g-2$ and the data of lepton universality in $\tau$ decays, but the mass splittings among extra Higgs bosons are strictly constrained.
\end{abstract}
%% \pacs{12.60.Fr, 14.80.Ec, 14.80.Bn}

\maketitle

%---------------------------------------------------------------------------
\section{Introduction}
The CDF collaboration presented their new result for the $W$-boson mass measurement \cite{cdfmw}
\bea
m_W=80.4335 \pm 0.0094 {\rm GeV}.
\eea 
The experimental central value has an
approximate $7\sigma$ discrepancy from the Standard Model (SM) prediction, $80.357 \pm 0.006$  GeV \cite{smmw}.
On the other hand, there has been a long-standing discrepancy between the SM prediction and experiment for the muon anomalous magnetic moment (muon $g-2$).
The combined result of the FNAL experiment \cite{fermig2} and the BNL experiment \cite{mug2-exp} has an approximate $4.2\sigma$ deviation from the SM prediction \cite{smg2-1,smg2-2,smg2-3}
\bea
\Delta a_\mu=a_\mu^{exp}-a_\mu^{SM}=(25.1\pm5.9)\times10^{-10}.
\eea 
Both the two deviations strongly imply existence of new physics beyond SM. 
Some plausible explanations have already performed for the CDF $W$-mass  \cite{2204.03693,2204.03796,2204.03996,2204.04183,2204.04191,2204.04202,2204.04204,2204.04286,2204.04356,2204.04514,2204.04559,2204.04770,2204.04805,
2204.04834,2204.05024,2204.05031,2204.05085,2204.05260,2204.05267,2204.05269,2204.05283,2204.05284,2204.05285,2204.05296,
2204.05302,2204.05303,2204.04688,2204.05728,2204.05760,2204.05942,2204.05962,2204.05965,2204.05975,2204.05992}.

Among various new physics models, two-Higgs-doublet models (2HDMs) are rather simple extensions of the SM (for a recent review, see, e.g., 
\cite{stu4}). The 2HDM introduces a second $SU(2)_L$
Higgs doublet and thus predicts two neutral CP-even Higgs bosons $h$ and $H$, one neutral pseudoscalar $A$ and a pair of charged Higgs boson $H^\pm$ \cite{t.lee}. The 2HDM can give additional corrections to the masses of gauge bosons via 
the self-energy diagrams exchanging extra Higgs fields. In addition, if the extra Higgs bosons have appropriate couplings to the leptons, the muon $g-2$ can be simply explained. Since both the $W$-mass and muon $g-2$ can be affected by the mass splittings among $H$, $A$, and $H^\pm$, in this note we take a 2HDM with the $\mu$-$\tau$ lepton flavor violation (LFV)
 interactions to study the possibility of a simultaneous explanation of both anomalies.  
In our analysis we will intensively examine the parameter space of this model by considering various relevant theoretical and experimental constraints. For the single explanation of muon $g-2$ using the Higgs doublet field with the $\mu$-$\tau$ LFV interactions, see e.g., 
 \cite{taumug2-3,0207302,10010434,150207824,151108544,151108880,1809.05857,1610.06587,160604408,190410908,1907.09845,1908.09845,2104.03242,
lfv11,lfv12,lfv13,1711.08430,2010.04266,2006.01934}.

The paper is organized as follows. In Sec. II we will introduce the 2HDM with the $\mu$-$\tau$ LFV interactions. 
In Sec. III and Sec. IV we 
study the W-boson mass and muon $g-2$ after imposing relevant theoretical and experimental constraints.
Finally, we give our conclusion in Sec. V.

\section{A two-Higgs-doublet model with $\mu$-$\tau$ LFV interactions}
The 2HDM with the $\mu$-$\tau$ LFV interactions may be derived from a  general 2HDM by take specific parameters.
Also it can be naturally obtained by introducing an inert Higgs doublet $\phi_2$ 
under a discrete $Z_4$ symmetry, and the $Z_4$ charge assignment is displaced in Table I \cite{190410908}.
\begin{table}
\caption{Assignment of $Z_4$ charge in the 2HDM with $\mu$-$\tau$-philic Higgs doublet.}
\label{tab:matter}
\centering
\begin{tabular}{cccccccccccc}
\hline
& ~$\phi_1$~ & ~$\phi_2$~ & ~$Q_L^{i}$~ & ~$U_R^i$~ & ~$D_R^i$~ & ~$L_L^e$~ & ~$L_L^\mu$~ & ~$L_L^\tau$~ & ~$e_R$~ & ~$\mu_R$~ & ~$\tau_R$~ \\ \hline
    ~~Z~$_4$~~~& $1$  & -1 &     1    & 1   &  1    &  1      & $i$      & $-i$     & 1   & $i$    & $-i$      \\ \hline
\end{tabular}
\end{table}
The Higgs potential with the $Z_4$ symmetry is given by
\begin{eqnarray} \label{V2HDM} \mathrm{V}   &=&   Y_1
(\phi_1^{\dagger} \phi_1) + Y_2 (\phi_2^{\dagger}
\phi_2)+ \frac{\lambda_1}{2}  (\phi_1^{\dagger}  \phi_1)^2 +
\frac{\lambda_2}{2}  (\phi_2^{\dagger} \phi_2)^2  \nonumber \\
&&+ \lambda_3
(\phi_1^{\dagger}  \phi_1)(\phi_2^{\dagger} \phi_2) + \lambda_4
(\phi_1^{\dagger} 
\phi_2)(\phi_2^{  \dagger} \phi_1)   + \left[\frac{\lambda_5}{2}   (\phi_1^{\dagger} \phi_2)^2 + \rm
h.c.\right].
\end{eqnarray}
Here all the parameters are real. Although $\lambda_5$ is the only possible complex parameter, it can be rendered real with a phase redefinition of one of the two Higgs fields. 
 The two Higgs doublets $\phi_1$ and $\phi_2$ are expressed by
\begin{equation} \label{field}
\phi_1= \left(\begin{array}{c}   G^+ \\
\frac{1}{ \sqrt{2}}\,(v+h+iG^0)
\end{array} \right)\,, \ \ \
\phi_2=\left( \begin{array}{c} H^+ \\
\frac{1}{\sqrt{2 }}\,(H+iA)
\end{array}\right) . \nonumber
\end{equation}
The $\phi_1$ field obtains a nonzero vacuum expectation value (VEV), $v$=246
GeV, while the $\phi_2$ field has zero VEV. The parameter $Y_1$ can be determined from the minimization condition for the Higgs potential, 
\beq
Y_1=-\frac{1}{2}\lambda_1 v^2.
\eeq
The fields $G^0$ and $G^+$ indicate Nambu-Goldstone bosons, which are eaten by the gauge bosons. 
The fields $A$ and $H^+$ represent the mass eigenstates of the CP-odd Higgs boson and
charged Higgs boson, whose masses are written as
\beq \label{masshp}
 m_{H^\pm}^2  =  Y_2+\frac{\lambda_3}{2} v^2, ~~~m_{A}^2  = m_{H^\pm}^2 +\frac{1}{2}(\lambda_4-\lambda_5) v^2.
 \eeq
There is no mixing between the two CP-even Higgs bosons $h$ and $H$, and their masses are   
\beq \label{massh}
 m_{h}^2  = \lambda_1  v^2\equiv   (125~{\rm GeV })^2, ~~~m_{H}^2   =  m_{A}^2+ \lambda_5 v^2.
\eeq

We obtain the masses of fermions via the Yukawa interactions with $\phi_1$,
 \beq \label{yukawacoupling} - {\cal L} = y_u\overline{Q}_L \,
\tilde{{ \phi}}_1 \,U_R + y_d\overline{Q}_L\,{\phi}_1 \, D_R +  y_\ell\overline{L}_L \, {\phi}_1
\, E_R + \mbox{h.c.}, \eeq
where $\widetilde{\phi}_1=i\tau_2 \phi^*_1$, and $E_R$, $U_R$ and $D_R$ stand for the three generations of right-handed fermion fields for the charged leptons, up-type quarks and down-type quarks.
 We define $L_L = (v_{L_i},\ell_{L_i})^{T}$ and $Q_L=(u_{L_i},d_{L_i})^T$  with $i$ representing generation indices. Under the $Z_4$ symmetry, the lepton Yukawa matrix $y_\ell$ is diagonal and therefore
the lepton fields ($L_L$, $E_R$) are mass eigenstates.

Under the $Z_4$ symmetry, the $\phi_2$ doublet is allowed to have $\mu$-$\tau$ interactions \cite{190410908},
\bea\label{lepyukawa2}
- {\cal L}_{LFV} & = &  \sqrt{2}~\rho_{\mu\tau}  \,\overline{L^\mu_{L}} \,  {\phi}_2
\,\tau_R  \, +  \sqrt{2}~\rho_{\tau\mu}\, \overline{L^\tau_{L}} \, {\phi}_2
\,\mu_R \, + \,  \mbox{h.c.}\,. \eea
 The additional Higgs bosons $H$, $A$ and $H^\pm$ only have $\mu$-$\tau$ LFV Yukawa couplings.
On the other hand, the SM-like Higgs boson $h$ has exactly same couplings to the gauge bosons and fermions as the SM Higgs, with no $\mu$-$\tau$ LFV couplings at the tree level. 

\section{The $S,~ T,~ U$ parameters and $W$-mass}
The model can produce main corrections to the masses of gauge bosons via 
self-energy diagrams exchanging extra Higgs fields. The oblique parameters $(S,~ T,~ U)$ \cite{stu,w-stu} represent radiative corrections to the two-point functions of gauge bosons.
Most effects on precision measurements can be described by these parameters. Recently,
 Ref. \cite{2204.03796} gave the values of these parameters from an analysis of precision electroweak data including the CDF new result of the $W$-mass,
\beq\label{fit-stu}
S=0.06\pm 0.10, ~~T=0.11\pm 0.12,~~U=0.14 \pm 0.09. 
\eeq
The correlation coefficients are given by
\beq
\rho_{ST} = 0.9, ~~\rho_{SU} = -0.59, ~~\rho_{TU} = -0.85.
\eeq
The $W$-boson mass can be inferred from the following relation \cite{w-stu},
\beq
\Delta m_W^2=\frac{\alpha c_W^2}{c_W^2-s_W^2}m_Z^2 (-\frac{1}{2}S+c_W^2T+\frac{c_W^2-s_W^2}{4s_W^2}U).
\eeq
In our analysis, we adopt $\textsf{2HDMC}$ \cite{2hc-1} to calculate the 2HDM corrections to $S,~T,~U$ parameters, and 
 perform a global fit to the predictions of $S,~T,~U$ parameters. As the global fit results are presented on two-dimension planes, a limit of $\chi^2 < \chi^2_{\rm min} + 6.18 $ is set to obtain 
 $2\sigma$ favored regions, where $\chi^2_{\rm min}$ is the minimum of $\chi^2$ corresponding the best fit point.
In addition, we consider theoretical constraints from perturbativity, vacuum stability and unitarity, which are described in detail in Appendix A.

We scan $m_H$, $m_A$ and $m_{H^\pm}$ parameters in the following ranges:
\begin{align}
&80 {\rm ~GeV} < m_{H^\pm} < 1000 {\rm ~GeV},~~ 65 {\rm ~GeV} < m_{A} < 1000 {\rm ~GeV},\nonumber\\
&10 {\rm  ~GeV} < m_{H} < 120 {\rm ~GeV},~~ 130 {\rm ~GeV} < m_{H} < 1000 {\rm ~GeV}.
\label{epot2}
\end{align}

%%%%%%%%%%%%%%%%%%%%
%%%%%%%%%%%%%%%%%%%%%
\begin{figure}[tb]
\centering
 \epsfig{file=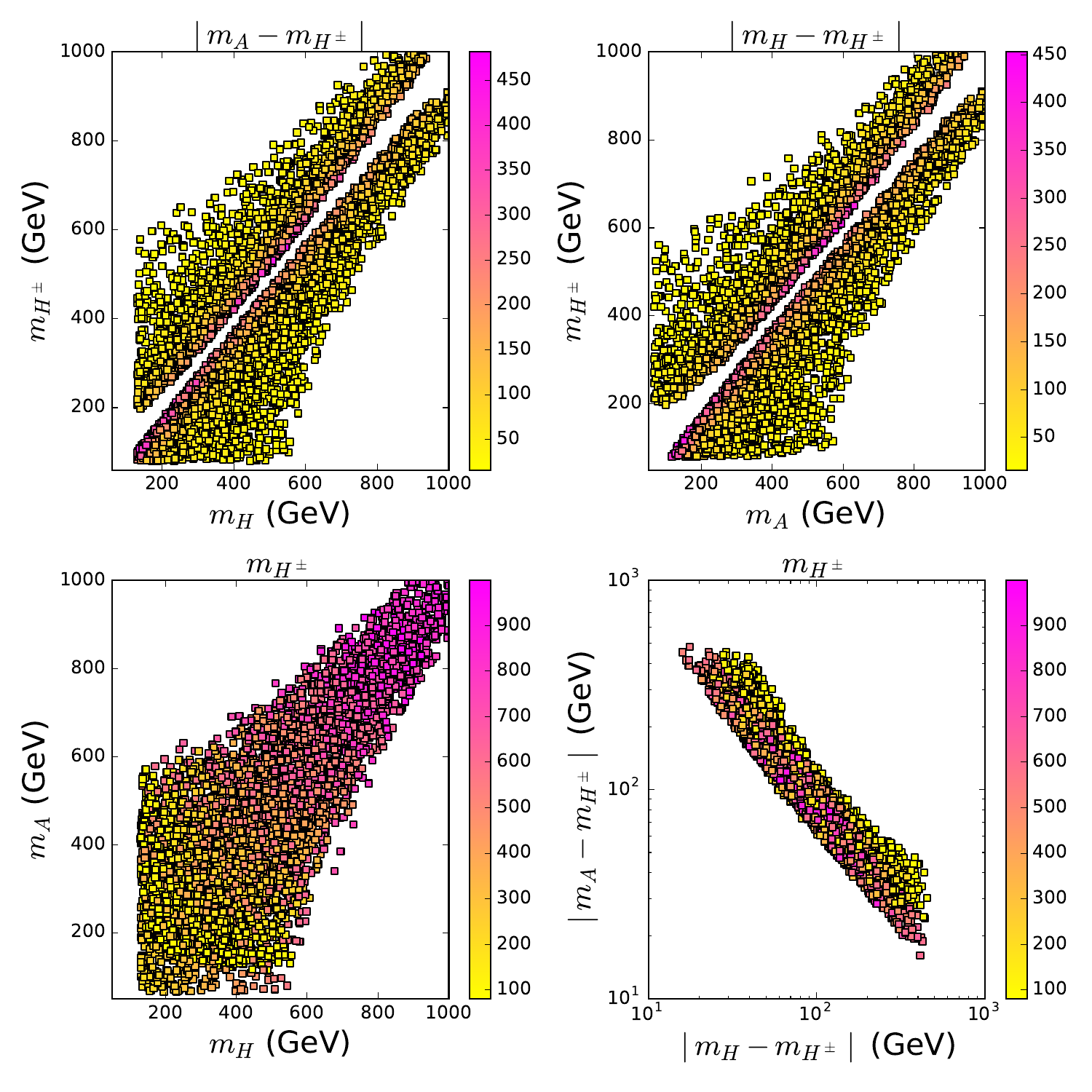,height=15cm}
 %\end{center}
\vspace{-0.2cm} \caption{For $m_H>$ 130 GeV, the samples explaining the CDF II results of the $W$-mass within $2\sigma$ range while satisfying
the constraints of the oblique parameters and theoretical constraints. The varying colors in each panel indicate 
the values of $\mid m_A-m_{H^\pm}\mid$, $\mid m_H-m_{H^\pm}\mid$ and $m_{H^\pm}$, respectively.} \label{stumw1}
\end{figure}
%%%%%%%%%%%%%%%%%%%%

%%%%%%%%%%%%%%%%%%%%%
\begin{figure}[tb]
\centering
 \epsfig{file=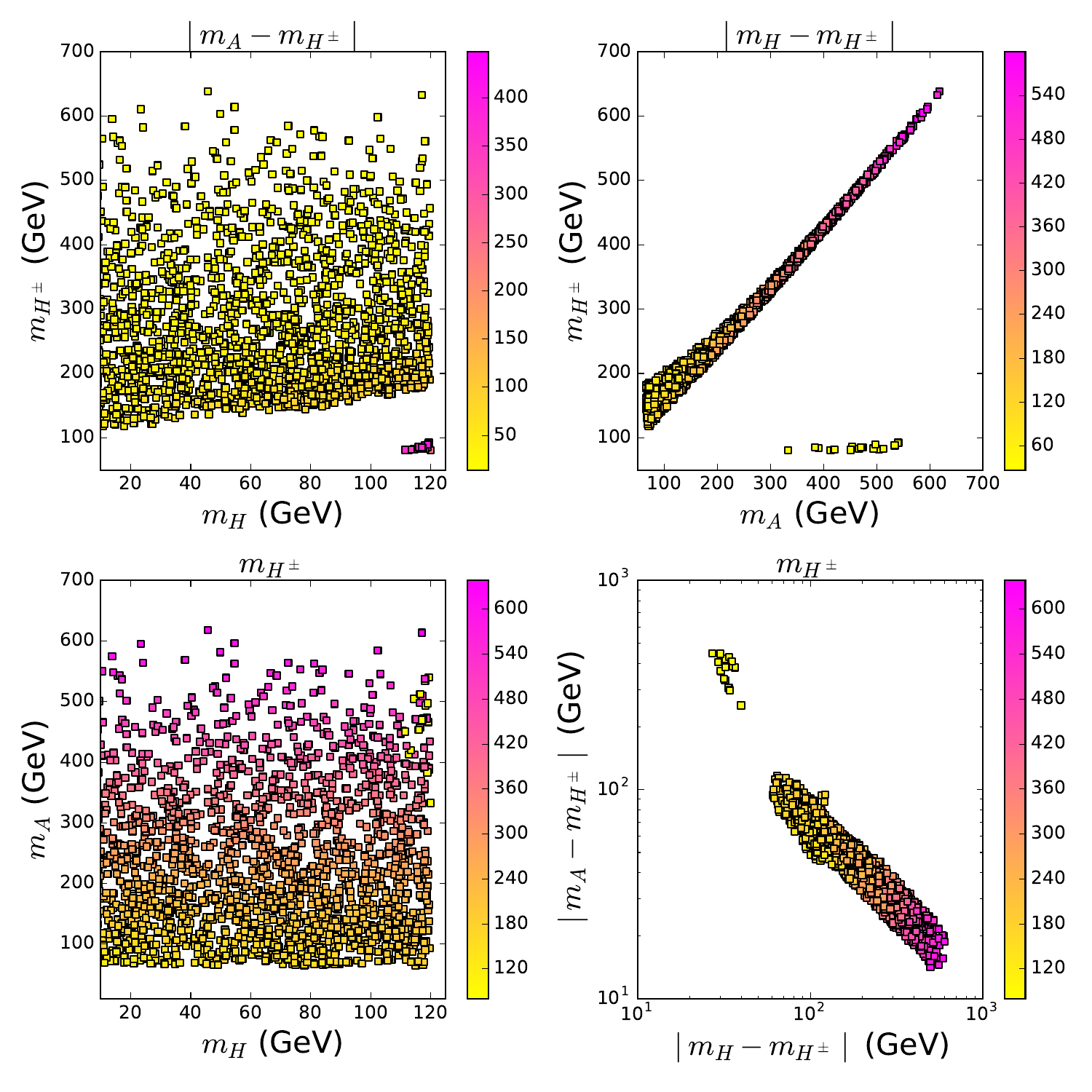,height=15cm}
 %\end{center}
\vspace{-0.2cm} \caption{Same as Fig. \ref{stumw1}, but for $m_H<$ 120 GeV.} \label{stumw2}
\end{figure}
%%%%%%%%%%%%%%%%%%%%

In Fig. \ref{stumw1} and Fig. \ref{stumw2}, we show the samples explaining the CDF $W$-boson mass measurement within $2\sigma$ range while satisfying
the constraints of the oblique parameters and theory. From Fig. \ref{stumw1} and Fig. \ref{stumw2}, we see that $H$ or $A$ is disfavored
 to exactly degenerate in mass with $H^\pm$, but their masses are allowed to be degenerate.
The mass splitting between $H^\pm$ and $H/A$ is imposed upper and lower bounds, and required to be approximately larger than 10 GeV.
When one of $m_H$ and $m_A$ is close to $m_{H^\pm}$, the other is allowed to have sizable deviation from $m_{H^\pm}$.
The $m_{H^\pm}$ and $m_{A}$ are favored to be smaller than 650 GeV for $m_H<120$ GeV (see Fig. \ref{stumw2}), and 
allowed to have more large values with increasing of $m_H$ (see Fig. \ref{stumw1}). 

Now we analyze the reason. In the model, the correction to the $T$ parameter is expressed as
\begin{eqnarray} \label{t-stu}
T\; & =&\;  \frac{1}{16\pi M_W^2s_W^2}\,  \Biggl\{ \biggl[
 \mathcal{F}(M_{H^\pm}^2,M_H^2) - \mathcal{F}(M_H^2,M_A^2) +  \mathcal{F}(M_{H^\pm}^2,M_A^2)  \biggr] \; \Biggr\}\, ,
\end{eqnarray}  
where the $\mathcal{F}$ function is \cite{stu1,stu2,stu3}
\beq\label{tfun}
\mathcal{F}(m_1^2,m_2^2) = \frac{1}{2}\, (m_1^2+m_2^2)-\frac{m_1^2m_2^2}{m_1^2-m_2^2}\;
\log{\left(\frac{m_1^2}{m_2^2}\right)}\, .
\eeq
The function $\frac{\mathcal{F}(m_1^2,m_2^2)}{m_W^2}$ and the factor $\frac{1}{16\pi s_W^2}$ in the $T$ parameter are usually larger than
those of $S$ and $U$ parameters. Therefore, in general one expects $T$ to be dominant in the oblique corrections. There are detailed discussions in 
Ref. \cite{tbig}.
The expressions in Eq. (\ref{t-stu}) and Eq. (\ref{tfun}) show that
the  $T$ parameter is sensitive to the mass splittings among $H,~A$ and $H^{\pm}$.
The $T$ parameter will be zero for $m_H=m_{H^\pm}$ or $m_A=m_{H^\pm}$, but takes a non-zero
value for $m_A=m_{H}$.
Therefore, the corrections of the model to the oblique parameters tend to become small 
as one of $m_H$ and $m_{A}$ approaches to $m_{H^\pm}$.
However, in order to accommodate the $W$-mass reported  by the CDF II collaboration,
the model needs to produce an appropriate value of $T$, which excludes $m_H=m_{H^\pm}$ or $m_A=m_{H^\pm}$. 

\section{Muon $g-2$, $\tau$ decays, and other relevant constraints}
The model gives additional corrections to the muon $g-2$ anomaly ($\Delta a_{\mu}$) via
 the one-loop diagrams involving the $\mu$-$\tau$ LFV couplings of $H$ and $A$ \cite{taumug2-3,0207302,10010434}, 
\bea
  \Delta a_{\mu} = \frac{m_\mu m_\tau  \rho^2}{8\pi^2}
  \left[\frac{ (\log\frac{m_H^2}{m_\tau^2} - \frac{3}{2})}{m_H^2}
  -\frac{\log( \frac{m_A^2}{m_\tau^2}-\frac{3}{2})}{m_A^2}
\right].
  \label{mua1}
\eea
Here we find $\Delta a_{\mu}>0$ for $m_A>m_H$.

Because the extra Higgs bosons have the $\mu$-$\tau$ LFV interactions, the model can 
affect the lepton flavor universality (LFU) in the $\tau$ lepton decays. The HFAG collaboration tests the LFU from ratios of the partial widths of a heavier lepton. They obtains \cite{tauexp}
\begin{align}
    \left( \frac{g_\tau}{g_\mu}\right) = 1.0011\pm0.0015, \left( \frac{g_\tau}{g_e}\right) = 1.0029\pm0.0015, \left( \frac{g_\mu}{g_e}\right) = 1.0018\pm0.0014, 
    \label{eq:hfag-data1}
\end{align}
using pure leptonic processes, namely
\begin{align}
    \left( \frac{g_\tau}{g_\mu}\right)^2 & \equiv \frac{
    \overline{\Gamma}(\tau\to e \nu \overline{\nu})}{\overline{\Gamma}(\mu \to e \nu \overline{\nu})},\\
    \left( \frac{g_\tau}{g_e}\right)^2 & \equiv \frac{
    \overline{\Gamma}(\tau\to \mu \nu \overline{\nu})}{\overline{\Gamma}(\mu \to e \nu \overline{\nu})},\\
    \left( \frac{g_\mu}{g_e}\right)^2 &\equiv \frac{
    \overline{\Gamma}(\tau\to \mu \nu \overline{\nu})}{\overline{\Gamma}(\tau \to e \nu \overline{\nu})},
\end{align}
with $\overline{\Gamma}$ representing the partial width which is normalized by the corresponding SM value.
$g_{e~(\mu,\tau)}$ denote the effective couplings between $e$ $(\mu,\tau)$ and $\nu_e$ $(\nu_\mu,\nu_\tau)$. 
With the two semi-hadronic processes,
\begin{align}
    \left( \frac{g_\tau}{g_\mu}\right)_h^2 & \equiv \frac{
    {\rm Br} (\tau\to h \nu) }{{\rm Br} (h\to \mu \overline{\nu})} \frac{2m_hm_\mu^2\tau_h}{(1+\delta_h) m_\tau^2 \tau_\tau} \left( \frac{1-m_\mu^2/m_h^2}{1-m_h^2/m_\tau^2} \right)^2,
\end{align}
where $h$ indicates $\pi$ or $K$, they measure 
\begin{align}
    \left( \frac{g_\tau}{g_\mu}\right)_\pi = 0.9963\pm0.0027, ~~~~~\left( \frac{g_\tau}{g_\mu}\right)_K = 0.9858\pm0.0071.
    \label{eq:hfag-data2}
\end{align}
The statistical correlation matrix for the five fitted coupling ratios is 
\begin{align}
    \left[
    \begin{array}{ccccc}
    1 & 53\% & -49\% & 24\% & 12\% \\
    53\%  & 1     &   48\% & 26\%    & 10\% \\   
    -49\%  & 48\%  & 1       &   2\% & -2\% \\
    24\%  & 26\%  & 2\%  &     1    &     5\% \\
    12\%  & 10\%  & -2\%  &  5\%  &   1 
\end{array} \right].
\label{eq:hfag-corr}
\end{align}

In this model, we can calculate
\begin{eqnarray} \label{tau-loop}
&&\bar{\Gamma}(\tau\to \mu \nu\bar{\nu})= (1+\delta_{\rm loop}^\tau)^2~(1+\delta_{\rm loop}^\mu)^2+\delta_{\rm tree},\nonumber\\
&&\bar{\Gamma}(\tau\to e \nu\bar{\nu})= (1+\delta_{\rm loop}^\tau)^2,\nonumber\\
&&\bar{\Gamma}(\mu\to e \nu\bar{\nu})= (1+\delta_{\rm loop}^\mu)^2.
\end{eqnarray}
The tree-level correction $\delta_{\rm tree}$ is from the contribution of $H^\pm$ to $\tau\to \mu\nu\overline{\nu}$, 
\beq
\delta_{\rm tree}=4\frac{m_W^4\rho^4}{g^4 m_{H^{\pm}}^4},
\label{delta-tree}
\eeq
which can give a positive correction. $\delta_{\rm loop}^\mu$ and $\delta_{\rm loop}^\tau$ are the corrections to vertices $W\bar{\nu_{\mu}}\mu$ and $W\bar{\nu_{\tau}}\tau$ from the one-loop diagrams containing $A$, $H$, and $H^\pm$, respectively. As we
assume $\rho_{\mu\tau}=\rho_{\tau\mu}$ in the lepton Yukawa matrix, the two corrections are identical \cite{190410908,tavv-1,mu2h16},
\beq
\delta_{\rm loop}^\tau=\delta_{\rm loop}^\mu={1 \over 16 \pi^2} {\rho^2} 
\left[1 + {1\over4} \left( H(x_A) +  H(x_H) \right)
\right]\,, 
\label{delta-loop}
\eeq
where $H(x_\phi) \equiv \ln(x_\phi) (1+x_\phi)/(1-x_\phi)$ with $x_\phi=m_\phi^2/m_{H^{\pm}}^2$.
Meanwhile, for the semi-hadronic processes, we have
\beq
\left( g_\tau \over g_\mu \right) =\left( g_\tau \over g_\mu \right)_K = \left( g_\tau \over g_\mu \right)_\pi.
\eeq

In our study, we perform a global fit to the predictions of these five ratios. 
Note that there is a vanishing eigenvalue in the covariance matrix constructed
from Eq. (\ref{eq:hfag-data1}), Eq. (\ref{eq:hfag-data2}) and Eq. (\ref{eq:hfag-corr}), and therefore such a degree of freedom is removed. With remaining four degrees of freedom, the $2\sigma$ confidence level region is obtained by adopting a limit of $\chi^2_\tau<9.72$. Thus, the surviving samples are much more consistent with the experimental results than the SM, which has a $\chi^2_\tau$ of $12.25$. 
Also the model can affect the LFU in the $Z$-boson decays \cite{zexp}, and the constraints from the $Z$-boson decays are generally weaker than those from the $\tau$ decays.

In the model, the fields $H$, $A$ and $H^{\pm}$ have no couplings to quarks, and therefore they are produced at the LHC mainly via
the electroweak processes, $pp\to W^{\pm *} \to H^\pm A/H$, $pp\to  Z^* \to HA$ and $pp\to Z^*/\gamma^* \to H^+H^-$.
The final state signal mainly includes multi-leptons, and therefore the 
multi-lepton event searches at the LHC can impose stringent constraints which require 
$m_H$ to be larger than 560 GeV \cite{1908.09845}. 
Also a very light $H$ may escape the constraints of the direct searches at the LHC \cite{2104.03242}.
In this case, the explanation of muon $g-2$ anomaly requires a very small $\rho$, which leads that the 
contributions of the model to the $\tau$ decays are too small to explain the data of LFU in
the $\tau$ decays. Therefore, in this work we discuss the scenario of $m_H>560$ GeV. In addition, to respect the perturbativity we choose the Yukawa coupling parameter $\rho<1$.

At the tree level, the 125 GeV Higgs has the same couplings to the SM particles as in the SM.
The $h\to \gamma\gamma$ decay will be corrected by the one-loop diagram of the charged Higgs \cite{1312.4759}. We consider the bound of the diphoton signal strength \cite{smmw}, 
\beq
\mu_{\gamma\gamma}= 1.11^{+0.1}_{-0.09}.
\eeq

%-----------------------
\begin{figure}[tb]
\centering
 \epsfig{file=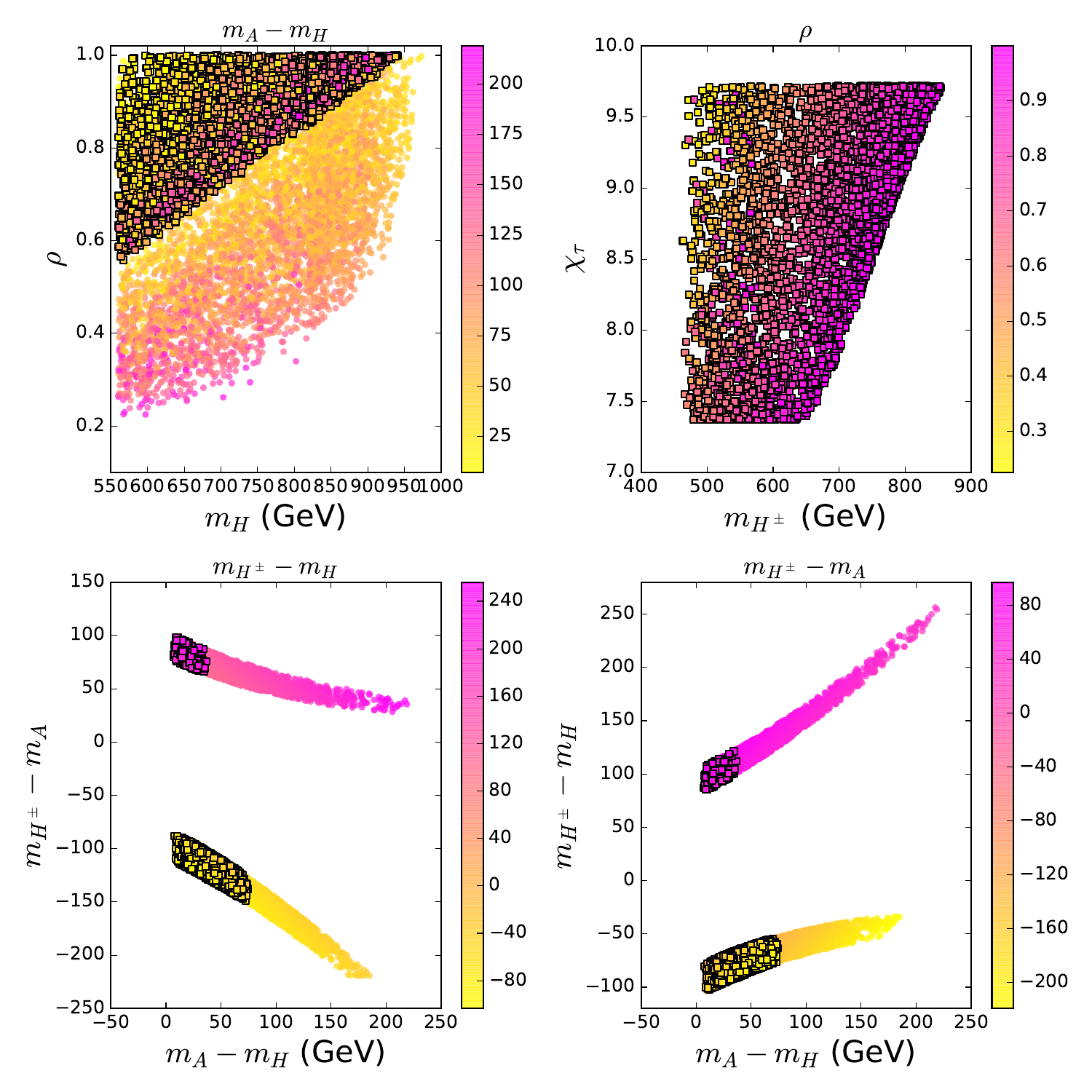,height=15cm}
 %\end{center}
\vspace{-0.5cm} \caption{The bullets can explain
the muon $g-2$ within $2\sigma$ range, while the squares can explain the muon $g-2$ and the $\tau$-decays within
$2\sigma$ ranges while satisfying the constraints of $Z$-decays.
Other relevant constraints from the theory, the oblique parameters, the CDF II $W$-mass and the diphoton signal data of the 125 GeV Higgs are also satisfied. The varying colors in each panel indicate 
the values of $m_A-m_{H}$, $\rho$, $m_{H^\pm}-m_H$ and $m_{H^\pm}-m_A$, respectively.} \label{g2ta}
\end{figure}
%%%%%%%%%%%%%%%%%%%%
In Fig. \ref{g2ta}, we project the surviving samples explaining 
the muon $g-2$ anomaly and the LFU in $\tau$-decays within $2\sigma$ ranges, with other relevant constraints from the theory,  the oblique parameters, the CDF II $W$-mass and the diphoton signal data of the 125 GeV Higgs being satisfied. 
Eq. (\ref{mua1}) shows that $H$ and $A$ respectively give positive and negative contributions to the muon $g-2$, and their contributions are suppressed by their masses. Therefore, the explanation of muon $g-2$ requires $m_A > m_H$ and $\rho$ to become large with increasing of $m_H$ and decreasing of $(m_A-m_H)$ (see the upper-left panel).

The ratio $\left( g_\tau \over g_e \right)$ in the $\tau$ decays has approximately $2\sigma$ deviation from the SM. Therefore, enhancing $\Gamma(\tau\to \mu \nu\bar{\nu})$ may give a better fit to the data of LFU in the $\tau$ decays.
The decay $\tau\to \mu\nu\nu$ obtains a positive correction from the $\delta_{\rm tree}$ term of Eq. (\ref{tau-loop}).
Such a term is from the tree-level diagram mediated by the charged Higgs, and proportional to
$\rho^4/m^2_{H^\pm}$. Therefore, the upper-right panel of Fig. \ref{g2ta} shows that 
the value of $\chi_\tau^2$ tends to become large 
with increasing of $m_{H^\pm}$ and decreasing of $\rho$. In addition, because of
the constraints of the oblique parameters and $W$-boson mass, the upper-left panel of
Fig. \ref{stumw1} shows that $m_{H^\pm}$ tends to increase with $m_H$, especially for a
large $m_H$, which implies that the value of $\chi_\tau^2$ becomes large with  
increasing of $m_H$. Therefore, the upper-left panel of Fig. \ref{g2ta} shows that
a simultaneous explanation of the muon $g-2$ and LFU in $\tau$ decays
favors a large $\rho$ which increases with $m_H$.

From the lower panel of Fig. \ref{g2ta}, we see that the mass splittings among 
$H$, $A$ and $H^\pm$ are stringently constrained in the region
 simultaneously explaining the $W$-mass, muon $g-2$ and LFU in $\tau$ decays, i.e., 
10 GeV $<m_A-m_H<$ 75 GeV, 65 GeV $<m_{H^\pm}-m_A<$ 100 GeV, 85 GeV $<m_{H^\pm}-m_H<$ 125 GeV
( -150 GeV $<m_{H^\pm}-m_A<$ -85 GeV, -105 GeV $<m_{H^\pm}-m_H<$ -55 GeV).

In the model, pair productions of extra Higgs bosons via electroweak processes at LHC lead to detectable multi-leptons signal containing $\mu$ and $\tau$. The current lower bound of 560 GeV on $m_H$ is mainly obtained from the CMS search for electroweak production of charginos and neutralinos in multilepton final states \cite{Sirunyan:2017lae}. By naively normalizing number of background and signal events, the lower bound on $m_H$ can be pushed to 700 GeV for 300 fb$^{-1}$ integrated luminosity data. Adopting the same procedure, we estimate that it needs more than 3000 fb$^{-1}$ integrated luminosity data to cover the whole surviving parameter, namely $m_H<950$ GeV. However, this CMS search is originally designed for searching SUSY particles. A dedicated study, focusing on $\mu\tau$ final states and improving signal regions for high mass region, can greatly reduce the required integrated luminosity to acceptable level, which is beyond the scope of this paper.

\section{Conclusion}
We examine the CDF II $W$-boson mass and the FNAL muon $g-2$ in the 2HDM 
in which the extra Higgs doublet has the $\mu$-$\tau$ LFV interactions. Imposing the 
theoretical constraints, we found that 
the CDF II $W$-boson mass disfavors $H$ or $A$ to degenerate in mass with $H^\pm$, and
the mass splitting between $H^\pm$ and $H/A$ is favored to be larger than 10 GeV.
The $m_{H^\pm}$ and $m_{A}$ are favored to be smaller than 650 GeV for $m_H<120$ GeV, and allowed to
have more large values with increasing of $m_H$. Considering other relevant experimental
constraints, we found that the mass splittings among $H,~A$ and $H^\pm$ are stringently restricted in the parameter space which can  simultaneously explain the CDF II $W$-mass, the FNAL muon $g-2$, and the data of LFU in the $\tau$ decays.

\section*{Acknowledgment}
We thank Lei Wu for helpful discussions.
This work was supported by the National Natural Science Foundation
of China under grants 11975013, 12105248, 11821505, 12075300, 12075213,
by Peng-Huan-Wu Theoretical Physics Innovation Center (12047503),
by the CAS Center for Excellence in Particle Physics (CCEPP), 
and by the Key Research Program of the Chinese Academy of Sciences, Grant NO. XDPB15. 

\begin{appendix}
\section{Theoretical constraints}
\subsection{Perturbativity}
The quartic couplings of the scalar potential in Eq. (\ref{V2HDM}) cannot be too large individually, for otherwise
the theory will no longer be perturbative. Thus we demand 
\beq
\mid\lambda_{1,2,3,4,5}\mid \leq 4\pi.
\eeq

\subsection{Vacuum stability}
Vacuum stability requires the potential to be bounded from below 
and stay positive for arbitrarily large values of the fields.
The requirement leads to restrictions on the parameters of the model,
\beq
\lambda_1 > 0,~~\lambda_2 > 0,~~\lambda_3 + \sqrt{\lambda_1\lambda_2} > 0,~~\lambda_3 + \lambda_4 - \mid\lambda_5\mid +\sqrt{\lambda_1\lambda_2} > 0.
\eeq

\subsection{Unitarity}
The amplitudes for scalar-scalar scattering $s_1 s_2 \to s_3 s_4$
at high energies respect unitarity, which leads to the following bounds on the parameters of the model \cite{unit-2h1,unit-2h2},
\begin{eqnarray} \label{unitarity}
|a_{\pm}|, |b_\pm|, |c_\pm|, |{\tt e}_\pm|, |{\tt f}_\pm|, |{\tt g}_\pm|
\,\le\, 8\pi \, ,
\end{eqnarray}
with 
\begin{eqnarray}
a_\pm^{} &=& \tfrac{3}{2}(\lambda_1+\lambda_2) \pm \sqrt{\tfrac{9}{4}(\lambda_1-\lambda_2)\raisebox{0.3pt}{$^2$}+(2\lambda_3+\lambda_4)^2} \,, \\
b_\pm^{} &=& \tfrac{1}{2}(\lambda_1+\lambda_2) \pm
\sqrt{\tfrac{1}{4}(\lambda_1-\lambda_2)\raisebox{0.3pt}{$^2$}+\lambda_4^2} \,, \\
c_\pm^{} \,&=&\, \tfrac{1}{2}(\lambda_1+\lambda_2) \pm
\sqrt{\tfrac{1}{4}(\lambda_1-\lambda_2)\raisebox{0.3pt}{$^2$}+\lambda_5^2} \,, \\
{\tt e}_\pm^{} &=& \lambda_3^{} + 2 \lambda_4^{} \pm 3 \lambda_5^{} \,, \\
{\tt f}_\pm^{} \,&=&\, \lambda_3^{} \pm \lambda_4^{} \,,\\
{\tt g}_\pm \,&=&\, \lambda_3^{} \pm \lambda_5^{} \,.
\end{eqnarray}

\end{appendix}

\end{document}